\documentclass[fleqn,twoside]{article}
\usepackage{espcrc2}
\usepackage{epsfig}
\usepackage{bm}

\usepackage{graphicx}
\usepackage[figuresright]{rotating}

\newcommand{\AmS}{{\protect\the\textfont2
  A\kern-.1667em\lower.5ex\hbox{M}\kern-.125emS}}

\title{QCD String Spectrum 2002}

\author{K. Jimmy Juge\address{Institute for Theoretical Physics, University 
   of Bern, Sidlerstrasse 5, CH-3012 Bern, Switzerland},
Julius Kuti\address{University of California at San Diego,
La Jolla, CA 92093, USA}
\thanks{Talk presented by J. Kuti.}
and	
Colin Morningstar\address{Carnegie Mellon University,
Pittsburgh, PA 15213, USA}}

\begin{document}

\begin{abstract}
Results from a 
comprehensive new analysis on the
excitation spectrum of the QCD string are presented.
A rapid onset of string formation is observed in the spectrum 
on a length scale of 2 fm, with Dirichlet boundary conditions.
The crossover from the short distance spectrum towards string
excitations and an observed fine structure in the 1--3 fm range
are related to effective string theory. The deficiencies
of the Nambu-Goto bosonic string model in describing the observed
spectrum are briefly discussed.
\vspace{1pc}
\end{abstract}

\maketitle

We report here our new comprehensive analysis of the rich low-lying energy 
spectrum of the excited gluon field
between a static quark--antiquark $(q\bar q)$ pair in the fundamental 
color representation of $SU(3)_c$.
For references on earlier work, we refer to our pilot journal
publication~\cite{JKM0}.
The analysis of each simulation
will be detailed in forthcoming publications. 
 
Three exact quantum numbers which are based on the symmetries of the problem
determine the classification scheme of the gluon excitation spectrum
in the presence of a static $q\bar q$ pair.
We adopt the standard notation from the physics of diatomic molecules
and use $\Lambda$ to denote the magnitude of the eigenvalue of the projection
${\bf J}_g\!\cdot\hat{\bf R}$ of the total angular momentum ${\bf J}_g$
of the gluon field onto the molecular axis with unit vector
$\hat{\bf R}$. The capital Greek
letters $\Sigma, \Pi, \Delta, \Phi, \dots$ are used to indicate states
with $\Lambda=0,1,2,3,\dots$, respectively.  The combined operations of
charge conjugation and spatial inversion about the midpoint between the
quark and the antiquark is also a symmetry and its eigenvalue is denoted by
$\eta_{CP}$.  States with $\eta_{CP}=1 (-1)$ are denoted
by the subscripts $g$ ($u$).  There is an additional label for the
$\Sigma$ states; $\Sigma$ states which
are even (odd) under a reflection in a plane containing the molecular
axis are denoted by a superscript $+$ $(-)$.  Hence, the low-lying
levels are labeled $\Sigma_g^+$, $\Sigma_g^-$, $\Sigma_u^+$, $\Sigma_u^-$,
$\Pi_g$, $\Pi_u$, $\Delta_g$, $\Delta_u$, and so on.  For convenience,
we use $\Gamma$ to denote these labels in general.

\begin{figure}[h]
\vskip -0.3in
\epsfxsize=3.0in
\epsfbox{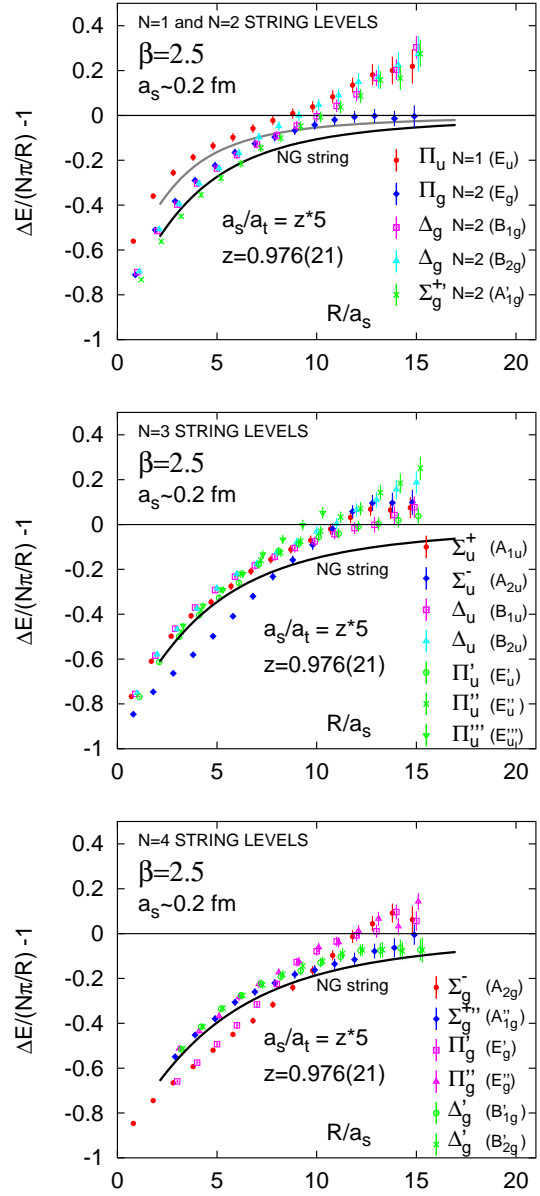}
\vskip -0.4in
\caption{
Energy gaps $\Delta E$ above $\Sigma^+_g$  are shown
in string units for quantum numbers
in continuum and lattice notation. 
The Nambu-Goto string is discussed in the text.
}
\label{fig:fig1}
\vskip -1.4in
\end{figure}

Restricted to the $R=0.2-3$ fm range of a selected simulation,
energy gaps $\Delta E$ above the ground state are compared to
asymptotic string gaps for 18 excited states 
in Fig.~\ref{fig:fig1}.
The quantity $\Delta E/(N\pi/R) - 1$ is plotted to show percentage deviations   
from the asymptotic string levels. 
The selected states correspond to $N=1,2,3,4$ string levels in the asymptotic
limit.  
The energy gaps, far below the null lines of 
the plots and strongly split at fixed $N$,
differ from the simple string gaps for $R<2$~fm.
The Nambu-Goto (NG) bosonic string does somewhat better. The spectrum
with fixed end boundary conditions in $D$ dimensions was first 
calculated in Ref.~\cite{arvis},
\begin{equation}
  E_N = \sigma R\biggl[ 1 - \frac{D-2}{12\sigma R^2}\pi+ \
  \frac{2\pi N}{\sigma R^2}\biggr]^{\frac{1}{2}} ~~,
\end{equation}
where $\sigma$ is the asymptotic string tension. The $D=4$ choice has
quantization problems and a tachyon
singularity in the spectrum at $R\approx 1/3$~fm.
The gaps $E_N$ are plotted in Fig.~\ref{fig:fig1} for all $N$
values (the gray line corresponds to $N=1$).
The states $\Sigma^{-}_u, \Sigma^{-}_g$ and $\Pi^{'}_g$ are in the worst
disagreement for $R<2$~fm. Above $2$~fm, several of the levels break away
from the NG formula and stay higher than the null level of the string gaps.
In addition, there is a pronounced fine structure at all
$R$ values. We now discuss three distinct regions in the spectrum.
%
%
\newline\noindent{\bf Short distance spectrum.} 
For $R \ll 1$~fm, the observed
level ordering is consistent with short distance physics
of gluon field excitations which are trapped around
the dominantly color octet static $q\bar q$ pair.  
A new approximate symmetry is expected in the
$R/\sqrt\sigma\ll 1$ limit since gluon field dynamics will only
depend softly on ${\bf R}$.
Gluon excitations will transform 
according to representations of $O(3)$ so that $\Lambda$ in the set $\Gamma$
will be replaced by the gluon angular momentum $L$.
Since this group is larger than $\Gamma$, several 
gluonic excitations between static quarks are 
expected to be approximately degenerate in the short-distance limit.
This approximate symmetry was 
first used within the short distance operator product
expansion of QCD in Ref.~\cite{soto}.
References to earlier work are given in~\cite{JKM0}.

\begin{figure}[t]
\epsfxsize=2.8in
\epsfysize=4.4in
\epsfbox{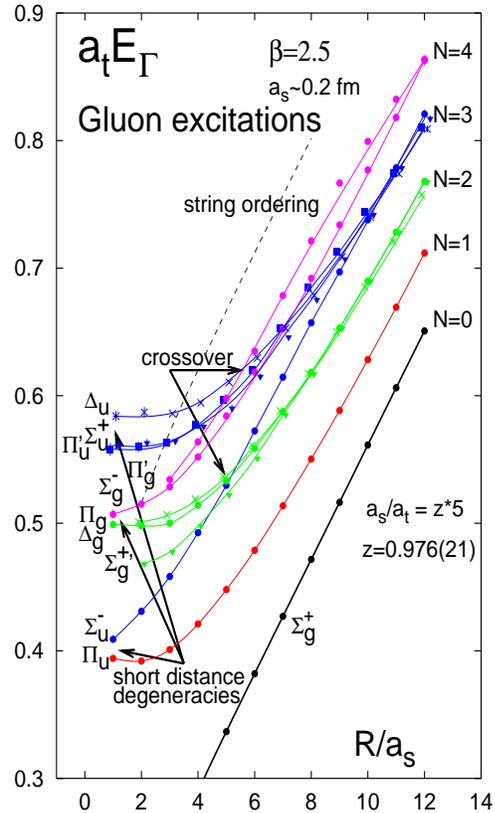}
\vskip -0.4in
\caption{Short distance degeneracies and crossover
in the spectrum. The solid curves are only shown for visualization.
The dashed line marks a lower bound for the onset of mixing effects
with glueball states which requires careful interpretation.
}
\label{fig:fig2}
\vskip -0.35in
\end{figure}

There exist ten states which are described 
by operators up to dimension three
with degenerate groups:
$(\Sigma^{+'}_g,\Pi_g)$, $(\Sigma^-_u,\Pi_u)$, 
$(\Sigma^-_g,\Pi^{'}_g,\Delta_g)$, and
$(\Sigma^+_u,\Pi^{'}_u, \Delta_u)$. 
Within each group one has different string quantum numbers mixed
together.
Fig.~\ref{fig:fig2} illustrates the remarkable 
working of the predicted degeneracies.  Only $\Delta_u$ and $\Sigma^{+'}_g$
show considerable soft breaking of the approximate symmetry
at the shortest $R$ values.
We note that the hierarchy of the doublet $(L=1)$ and triplet $(L=2)$ groups
follows the dimensional hierarchy of the operators, similar to
the dimensional hierarchy of glueball operators in Ref.~\cite{JK1}.
\newline\noindent {\bf Crossover region and Casimir energy.}
On the intermediate scale of $0.5$ fm $ < R < 2$~fm,
a rapid crossover of the energy levels towards
a string-like spectrum is prominent.
An interesting feature of this region is 
the rather accurate description of the $\Sigma_{\rm g}^+$ 
ground state energy by the empirical function
$E_0(R) = a + \sigma R - c\pi/12R$, 
with the fitted constant $c$ 
close to unity, once $R$ exceeds $0.5$~fm
(the Casimir energy of a thin flux line was calculated 
in Refs.~\cite{Luescher1,Luescher2}, yielding $c=1$).
This approximate agreement
is often interpreted as evidence for string formation.
The conformal charge $c$ and the spectrum are related. 
Our excitation spectrum, even the qualitative ordering
of the energy levels, does not agree well with the asymptotic bosonic string spectrum for $R < 1$~fm. 
A recent high precision calculation shows the rapid approach of $c_{\rm eff}(R)$ to the asymptotic Casimir value in the same $R$ range~\cite{Luescher3}. 
\newpage\noindent
It remains a challenge to explain the precocious onset of the asymptotic value of the conformal charge.
\par\noindent{\bf String limit}.
The rapid rearrangement of the energy levels 
towards asymptotic string-like ordering around $R \approx 2$~fm is remarkable. 
For example, the states $ \Sigma^-_u$ and
$ \Sigma^-_g $ break away from their respective short distance non-string
degeneracies to approach  string ordering for $R > 2$~fm separation.
After the rapid transition,
the level orderings and approximate degeneracies
at large $R$ match, without exception, those expected
of the asymptotic string modes.
However, the
precise separation $\pi/R$  of the energy levels
is not observed in the spectrum. Some of the expected
degeneracies are also significantly broken.

An explanation for this fine structure is expected to come 
from effective QCD string theory which is formulated in terms of the
Lagrangian of the two-dimensional position vector $\bm{\xi}$ of the string
defined in world sheet coordinates~\cite{Luescher1,Luescher2}.
Symmetries of the effective
Lagrangian require a derivative expansion of the form 
$ {\mathcal L}_{\rm eff} = \frac{1}{2}\alpha\partial_\mu\bm{\xi}\cdot
\partial_\mu\bm{\xi} + ...,$ where 
the dots represent terms with four, or more derivatives
and $\alpha$ is a dimensional constant.
For low frequency excitations, the first term, which describes
massless Goldstone modes (asymptotic string modes), dominates and higher
dimensional operators are expected to generate perturbative fine structure in powers of $R^{-1}$.

The Nambu-Goto action when expanded,
\begin{equation}
 {\mathcal L}_{\rm eff} = \alpha \sqrt{ {\rm 1}+\partial_\mu\bm{\xi}\cdot
\partial_\mu\bm{\xi} } =  
 \frac{1}{2}\alpha\partial_\mu\bm{\xi}\cdot
\partial_\mu\bm{\xi} + ...,
\end{equation}
generates higher dimensional operators with preset coefficients
which cannot describe rigidity, torsion, and other physical properties of the confining flux, if they exist.
Deviations of our spectrum from the NG prediction demonstrates 
that the coefficients in the effective QCD string action will differ 
significantly from those of the NG Lagrangian
(a particularly interesting form was proposed in Ref.~\cite{polchinski}).

This work was supported by the U.S.~DOE, Grant No. DE-FG03-97ER40546, the U.S. National Science Foundation under Award PHY-0099450 and the European
Community's Human Potential Programme, HPRN-CT-2000-00145.

\end{document}